# High-Efficiency Field-Free Superconducting Diodes in $Mo_2N$/FePt Bilayers

Juan S. Cruz,[1] Emilio De Biasi,[1,2] Juan Zapata,[1,2] Yeonkyu Lee,[3] Jeehoon Kim,[3*] Nestor Haberkorn[1,2#]

[1] Instituto Balseiro, Universidad Nacional de Cuyo, and Comisión Nacional de Energía Atómica, Av. Bustillo 9500, 8400 San Carlos de Bariloche, Argentina

[2] Instituto de Nanociencia y Nanotecnología (CNEA-CONICET), Nodo Bariloche, Av. Bustillo 9500, 8400 San Carlos de Bariloche, Río Negro, Argentina

[3] Department of Physics, Pohang University of Science and Technology, Pohang 37673, South Korea

Keywords: superconducting diodes, superconductor/ferromagnet hybrids, vortex dynamics, stripe domains, magnetic stray fields

Near-ideal, field-free superconducting diode behavior is demonstrated in room-temperature-grown $Mo_2N$/FePt bilayers. The response is consistent with remanent FePt stripe domains producing polarity-dependent thresholds for flux entry and vortex motion. Micromagnetic simulations reveal a periodic stray-field modulation across the stripes and asymmetric field enhancement near their ends at the bridge edges. One current polarity remains nearly dissipationless up to a sizable fraction of the estimated depairing current, whereas the opposite polarity develops dissipation at much lower current, yielding diode efficiencies up to approximately 85% in the thinner bilayer. The nonreciprocity can be continuously tuned by the in-plane magnetic preparation angle and reversibly switched by reverse-field excursions through the hysteretic reversal of the FePt layer. These results support edge-controlled flux entry and vortex motion as a route to high-efficiency superconducting diodes.

E-mail: jeehoon@postech.ac.kr (Jeehoon Kim); nestor.haberkorn@ib.edu.ar (Nestor Haberkorn)

## 1. Introduction

Controlling vortex motion in type-II superconductors is essential not only for limiting dissipation, but also for engineering nonreciprocal transport functionalities.[1,2] Because the resistive response is governed by vortex nucleation, entry, and motion under applied current, tailoring the vortex energy landscape through geometry, pinning, or magnetic textures provides a direct route to polarity-dependent transport. Within this framework, superconducting diodes can be understood as a manifestation of asymmetric vortex dynamics, leading to different critical currents for opposite bias directions. Such nonreciprocity can also arise from other physical mechanisms, including electronic effects in tunnel-junction and van der Waals systems.[3,4] In superconductor/ferromagnet (SC/FM) hybrids, however, magnetic stray fields (MSF) provide a particularly effective way to control flux entry and depinning, making these systems a promising platform for realizing superconducting diodes through magnetic control of vortex motion.[5,6,7]

For device applications, superconducting diode performance is typically evaluated in terms of field-free operation, the diode efficiency $\eta = \left||I_c^+| - |I_c^-|\right| / \left(|I_c^+| + |I_c^-|\right)$, and the accessible current window over which the nonreciprocal response can be tuned.[8] Here, $I_c^+$ and $I_c^-$ denote the positive magnitudes of the critical currents for opposite bias directions. The efficiency quantifies the magnitude of the nonreciprocity, while the branch with the larger $|I_c|$ defines the diode polarity. Within SC/FM hybrids, early studies showed that remanent magnetization can reshape the edge stray-field profile and generate polarity-dependent critical currents through its interplay with the transport self-field, with critical-current asymmetries of about 10% reported in oxide bilayers.[9] Co-capped Nb microbridges in cross-line geometries subsequently exhibited switching-current asymmetries corresponding to $\eta \approx 33\%$ after magnetization and approximately 29% without prior magnetization, when expressed using the normalization defined above.[10] Py/Nb bilayers subsequently demonstrated zero-field bistable diode behavior,[11] governed primarily by asymmetric depinning at the sample edges.[6] More recently, EuS-based SC/FM bilayers reached $\eta = 65\%$ under a small applied field and retained approximately 21% at zero external field,[7] while planar NbN/AlN/Py microbridges demonstrated field-free, memory-programmable diode functionality with efficiencies up to approximately 40%.[12] More recently, Pt/Co/Nb heterostructures combining perpendicular magnetization with intentionally engineered geometric asymmetry achieved field-free efficiencies of approximately 54%.[13] These results motivate the search for still larger nonreciprocity in structurally simple SC/FM hybrids, without intentionally patterned geometric asymmetry.

Here, we investigate the superconducting diode behavior of $Mo_2N$/FePt bilayers, focusing on how magnetic stray fields from stripe domains control vortex dynamics. Room-temperature-grown $Mo_2N$ remains superconducting up to 8 K and preserves superconductivity down to thicknesses of only a few nanometers.[14] FePt, in turn, develops stripe-domain structures already above 40 nm when grown at room temperature,[15] far below the thickness typically required to stabilize comparable domain patterns in Py (>150–200 nm).[16] Moreover, its larger saturation magnetization (≈ 1100 emu/cm$^3$, compared with ≈ 700 emu/cm$^3$ in Py) is expected to favor stronger edge stray fields for similar film thicknesses. Together, these features make FePt a particularly attractive magnetic layer for enhancing vortex asymmetry in simple superconducting hybrids. Using a structurally simple bilayer platform fabricated entirely at room temperature, we demonstrate field-free diode efficiencies up to approximately 85% in disordered $Mo_2N$ thin films capped with 56 nm-thick FePt, where the nonreciprocal response is governed by the remanent magnetic configuration and strongly depends on the in-plane preparation-field direction relative to the applied current.

## 2. Results and Discussion

**Figure 1**(a) shows the normalized resistance as a function of temperature for the two bilayers. The superconducting critical temperatures, defined at 50% of the resistive transition, are approximately 4.5 K and 6.5 K for $Mo_2N$(10)/FePt and $Mo_2N$(20)/FePt, respectively. These values are slightly lower than those observed in the corresponding single $Mo_2N$ films,[14] likely due to a small reduction in the effective superconducting thickness caused by proximity effects.[18] The normal-state resistance of the bilayers is largely determined by the FePt layer, whose resistivity just above $T_c$ (≈ 70 μΩ·cm) is lower than that of $Mo_2N$ (≈ 120 μΩ·cm).[14,17]

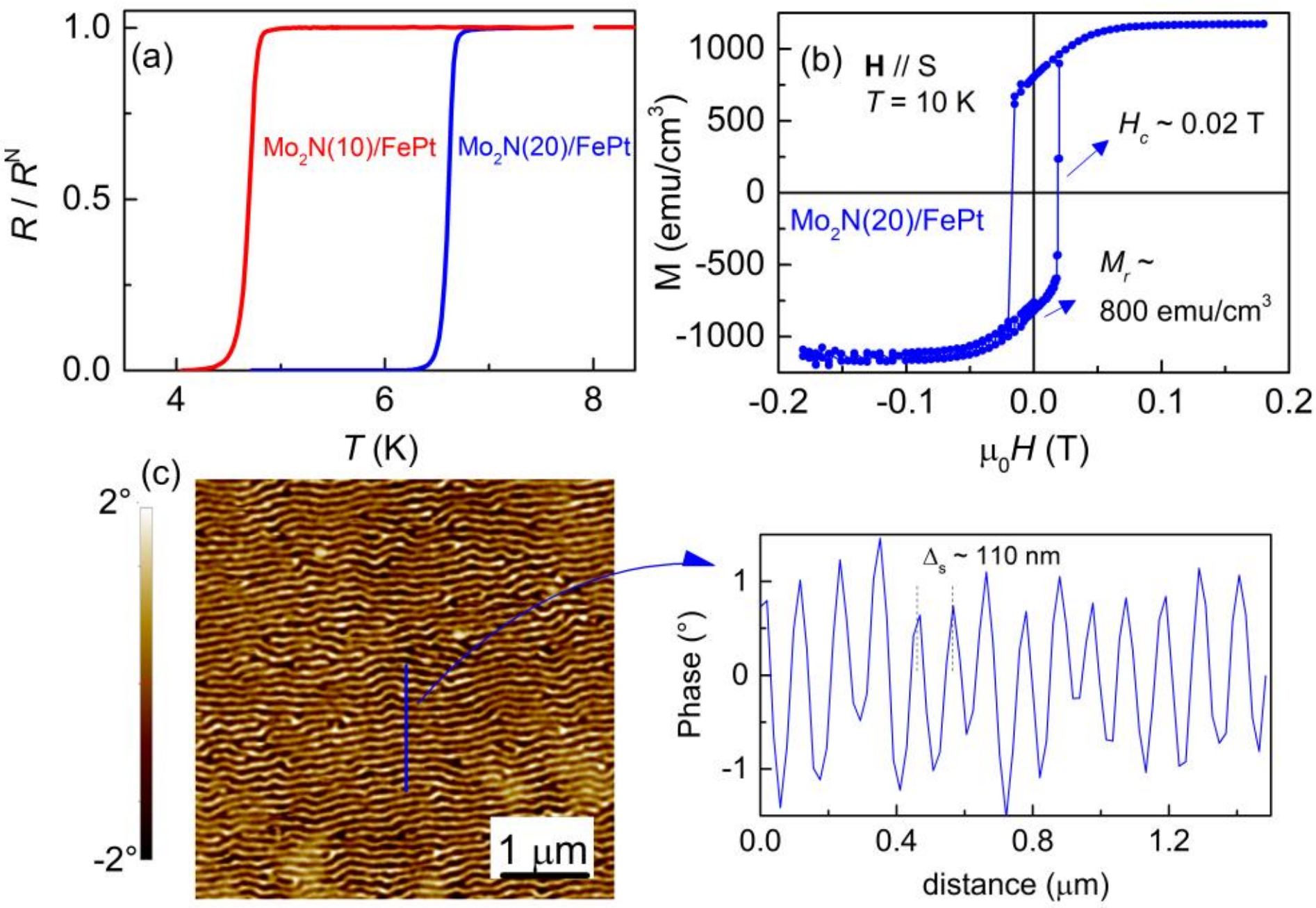

**Figure 1.** Basic superconducting and magnetic properties of the $Mo_2N$/FePt bilayers. (a) Normalized resistance as a function of temperature for $Mo_2N$(10)/FePt and $Mo_2N$(20)/FePt. (b) In-plane magnetization loop measured at 10 K for $Mo_2N$(20)/FePt. (c) Room-temperature MFM phase image showing FePt stripe domains; the line profile gives a characteristic periodicity of approximately 110 nm.

**Figure** 1(b) shows the in-plane magnetization loop of $Mo_2N$(20)/FePt at 10 K. It gives $M_s \approx$ 1100 emu/cm$^3$, $M_r \approx$ 800 emu/cm$^3$, and a coercive field of approximately 20 mT. The nearly linear approach to saturation is characteristic of the stripe-domain state.[15,17] The room-temperature MFM image independently confirms nanostripes with a periodicity $\Delta_s \approx$ 110 nm [**Figure** 1(c)]. MuMax3 simulations quantify this magnetic landscape (**Figure** 2).[19]

Using MuMax3 version 3.10, a 56-nm-thick strip was modeled with finite edges along x and periodic boundary conditions along y; for stripes perpendicular to the transport current, these axes correspond to the bridge width and current direction, respectively. The parameters $M_s$ = $1.08 \times 10^6$ A/m and $K_{u1} = 1.9 \times 10^5$ J/m$^3$, with the anisotropy axis along *z*, were chosen consistently with the magnetization measurements and previous FePt stripe-domain studies.[15,17] They give a quality factor $Q = \frac{2K_{u1}}{\mu_0 {M_s}^2} \approx 0.26$, consistent with a weak-stripe state. The remanent magnetization was used to reconstruct the perpendicular stray field on a plane 10 nm below the FePt layer. The mesh, exchange stiffness, relaxation protocol, and field calculation are detailed in the Supporting Information.

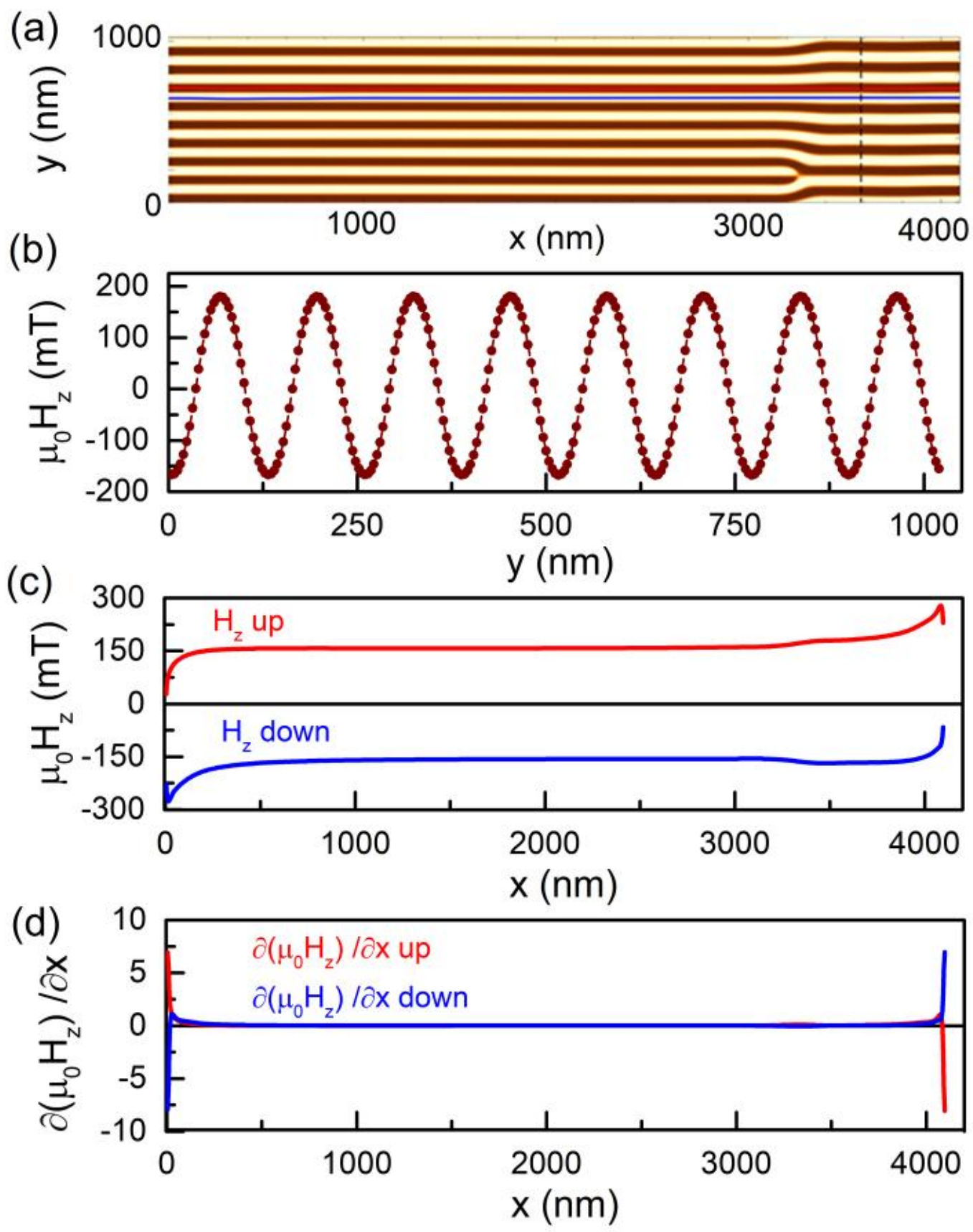

**Figure 2.** Micromagnetic configuration and calculated FePt stray field. (a) Normalized perpendicular magnetization $m_z$ at mid-thickness of the 4096 × 1024 × 56 $nm^3$ strip. Red and blue lines mark paths through opposite stripe polarities; the dashed line marks the transverse cut. (b) Transverse field profile, with period $\Delta_s$ = 128 nm and amplitude approximately 180 mT. (c) Field along the red and blue paths, showing opposite signs and asymmetric edge enhancement. (d) Corresponding longitudinal derivatives. All field profiles are evaluated 10 nm below the lower FePt surface. In this exterior region, $B_z = \mu_0 H_z$; values are reported in mT.

**Figure 2**(a) shows the relaxed stripe pattern, aligned along x and modulated along y, with a period $\Delta_s$ = 128 nm. This period is of the same nanoscale as the approximately 110 nm measured by MFM, although the finite periodic simulation cell selects eight periods within its 1.024 μm extent along y. The transverse cut in **Figure** 2(b) gives a nearly sinusoidal $B_z$ modulation with an amplitude of approximately 180 mT. Along the two paths parallel to the stripes, $\mu_0 H_z$ retains opposite signs and approaches approximately +160 and −160 mT away from the edges and the region of stripe distortions [**Figure** 2(c)]. The positive-field path is enhanced near the right edge and reduced near the left edge, while the negative-field path shows the complementary behavior, with its magnitude enhanced near the left edge and reduced near the right edge, with extrema approaching 280 mT in magnitude. The strongest longitudinal gradients, reaching approximately 8 mT/nm, are concentrated near these edges

[**Figure** 2(d)]; weaker structure remains near the stripe distortions, whereas the gradients are small in the regular interior. These profiles distinguish the transverse field modulation associated with the stripe texture from the longitudinal edge corrections that can affect vortex entry and motion. The calculated stray-field landscape provides a plausible route to polarity-dependent flux entry; however, the present magnetostatic model does not resolve the vortex distribution or dynamics.

**Figure 3** summarizes the physical picture proposed for the remanent superconducting diode state. Panel (a) shows the device and the three remanent stripe-domain configurations studied here. We define +*x* from the left to the right edge, +*y* along the positive-current direction, and +*z* from $Mo_2N$ toward FePt. The micromagnetic configuration displayed in **Figure** 2 corresponds to the remanent orientation of configuration I in **Figure** 3(b), with the magnetization **M** along -*x*. The macroscopic edge-field component is superimposed on the alternating local stripe fields, enhancing the negative $B_z$ extrema near the left edge and the positive $B_z$ extrema near the right edge [Figure 2(c)]. Within the proposed edge-entry regime, the sign of the strongest local $B_z$ at each edge selects the vorticity of the injected flux: antivortices at the left edge and vortices at the right edge. The vortex and antivortex symbols in **Figure** 3(c) represent this edge-selected vorticity. The calculated edge enhancement, with local $|B_z|$ maxima approaching 280 mT and longitudinal gradients reaching approximately 8 mT/nm, can favor nucleation near the stripe terminations and provide magnetic confinement of the resulting flux near the edges. A spontaneous interior vortex–antivortex population cannot be inferred from the field amplitude alone: its formation also depends on the magnetic texture and superconducting dimensions, as discussed below.[20] In our 5 μm-wide bridges, the transport self-field estimated from $B_{s-f} = \left(\frac{\mu_0 I}{2\pi w}\right) \ln(\frac{2w}{d})$ is of the order of 0.2 mT/mA,[21] where *d* is the superconducting-layer thickness. Over the relevant current range, it remains approximately 1 mT or less and is therefore much smaller than the calculated FePt stray fields.

Within the proposed edge-entry regime, consistent with the polarity-dependent transport behavior discussed below, dissipation is controlled by the competition between magnetic confinement near the edges and the current-induced Lorentz force. As discussed by Carapella *et al*,[6] this mechanism does not require a conventional periodic vortex-ratchet potential. For configuration I, negative current drives the edge-selected antivortices at the left edge and vortices at the right edge toward the interior [**Figure** 3(c)]. Once the magnetic confinement is overcome, guided motion along the stripe-defined channels can produce dissipation at relatively low current. For positive current, the Lorentz force points outward and reinforces the confinement against inward penetration, suppressing flux injection into the interior. This response can favor a vortex-depleted transport region and is consistent with the nearly dissipationless branch preceding abrupt switching. A qualitative precedent for current-polarity-dependent vortex-depleted channels is provided by SC/FM hybrids operating under field-compensation conditions.[22] Reversing the remanent magnetization reverses the edge-field asymmetry and hence the diode polarity. Within this proposed picture, rotating the remanent stripe configuration toward configuration III changes the relation between the magnetic channels and the Lorentz force, consistent with the observed suppression of the nonreciprocal response.

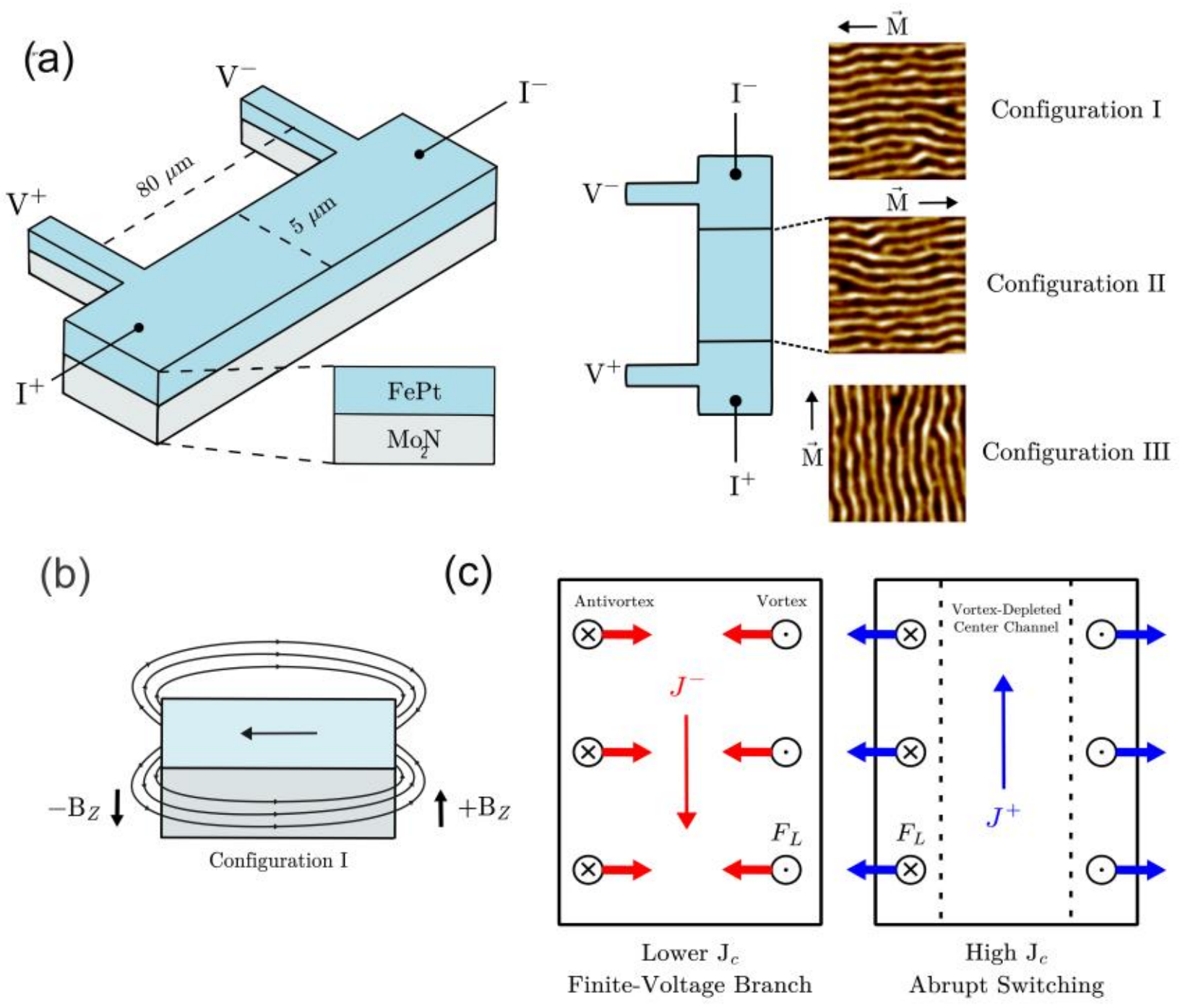

**Figure 3.** Proposed mechanism of the remanent superconducting diode response. (a) $Mo_2N$/FePt bridge and remanent configurations. Configurations I and II have the same stripe orientation and opposite remanent magnetization; configuration III is rotated by 90°. (b) Macroscopic edge-field component associated with configuration I, superimposed on the local stripe modulation. (c) Proposed edge-entry regime: the Lorentz force drives edge-nucleated vortices and antivortices inward for the dissipative polarity and outward for the robust polarity, favoring a vortex-depleted central region.

**Figure 4** summarizes the superconducting diode response in magnetic configuration I, as defined in **Figure** 3(a). All I–V curves were measured at zero applied magnetic field after preparing the remanent state. The $Mo_2N$(10)/FePt sample exhibits a strongly asymmetric I-V response [**Figure** 4(a)]: positive current produces an almost dissipationless branch up to an abrupt switching event, whereas negative current produces a finite-voltage branch at much lower current. The low-voltage |V| versus |I| characteristics in **Figure** 4(b) and 4(c) resolve the polarity-dependent critical currents. Their temperature dependences [**Figure** 4(d) and 4(e)] show that $I_c^+$ exceeds $I_c^-$ in both bilayers, with a larger relative separation in the 10-nm sample. The critical currents span tens of microamperes to approximately 4 mA across the two samples. At 2.8 K, the diode efficiency is approximately 78% for $Mo_2N$(10)/FePt and 50% for $Mo_2N$(20)/FePt. It increases with temperature, reaching approximately 85% at 3.4 K for the thinner bilayer and approximately 83% near 6 K for the thicker bilayer [**Figure** 4(f)]. At the same absolute temperature, the thinner bilayer exhibits a substantially larger efficiency.

The contrast between the abrupt positive-current branch and the low-threshold negative-current branch is consistent with the proposed edge-entry picture of **Figure** 3. The negligible voltage preceding positive-current switching, despite the weak longitudinal field variation that favors guided vortex propagation, strongly suggests that a sizable mobile vortex population is absent from the central transport region. This behavior is compatible with magnetic confinement opposing inward flux penetration and, within the proposed regime, with a vortex-depleted transport region, as discussed below. Conversely, the low dissipation threshold of the negative-current branch is consistent with guided vortex and antivortex motion along the stripe-defined magnetic channels once the edge confinement is overcome. The transverse stray-field modulation can confine flux to these channels, while the weak longitudinal field variation in the regular interior may facilitate propagation along them. Such a distinction between transverse confinement and relatively easy longitudinal motion is supported by Nb films decorated with Co nanostripes, where vortex guiding and strongly anisotropic depinning thresholds have been demonstrated.[23] In our bilayers, this landscape provides a plausible route to low-threshold dissipative transport without requiring uniformly weak pinning throughout the bridge. Thermally assisted overcoming of edge-entry or propagation barriers may contribute to the increase of $\eta$ with temperature, although the transport data do not distinguish their relative contributions. The particularly small $I_c^-$ in the thinner bilayer is consistent with this low-threshold scenario; attributing the difference between bilayers to weaker pinning additionally requires accounting for their different superconducting cross sections and reduced temperatures. These results motivate exploring whether weaker longitudinal pinning can further increase $\eta$ while preserving the high-current, nearly dissipationless branch protected by remanent edge confinement.

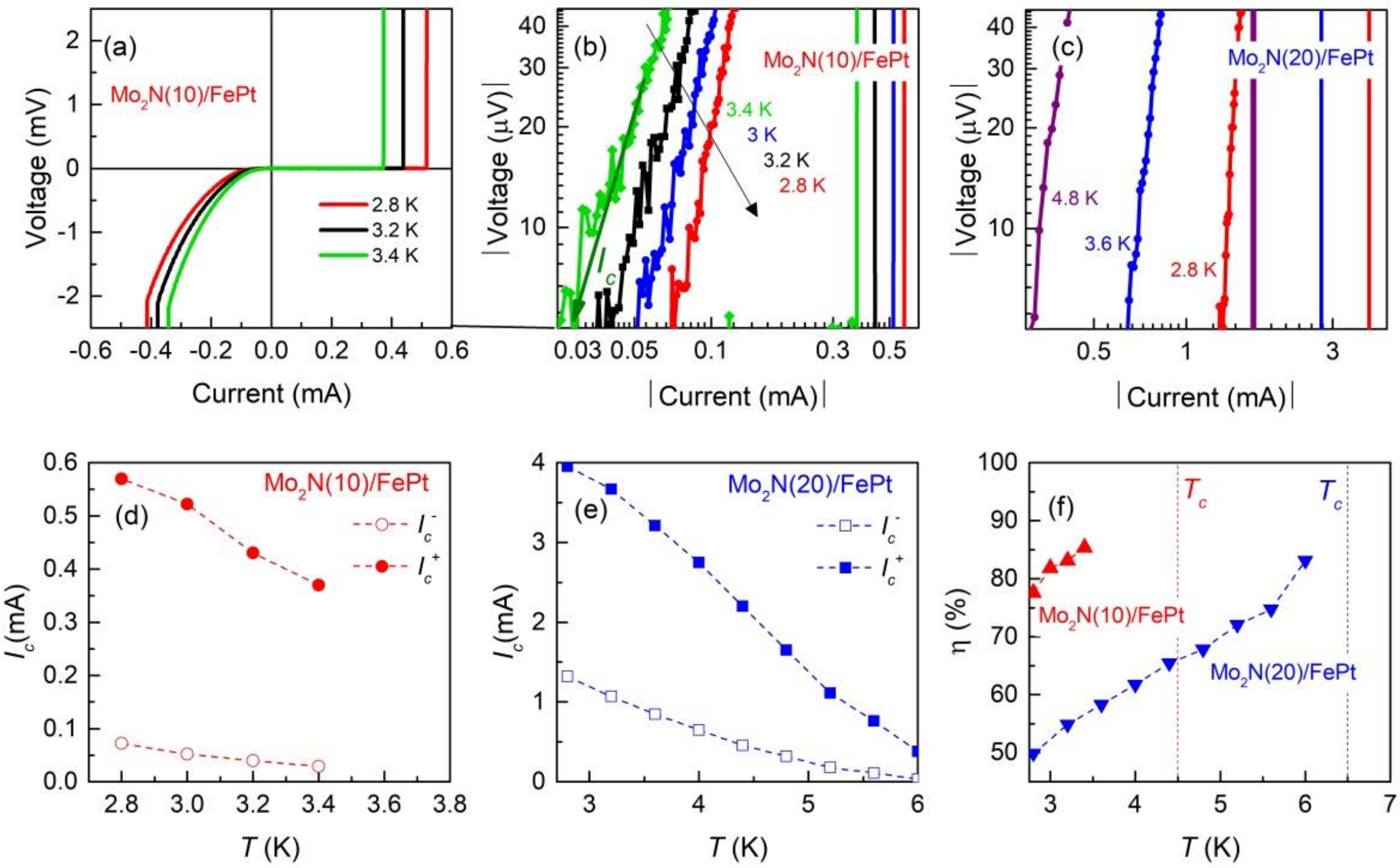

**Figure 4.** Superconducting diode response in magnetic configuration I of **Figure** 3. (a) I-V curves for $Mo_2N$(10)/FePt at selected temperatures. (b,c) Low-voltage |V| versus |I| characteristics on logarithmic scales for $Mo_2N$(10)/FePt and $Mo_2N$(20)/FePt, respectively. The low-current branches correspond to $I_c^-$; the vertical lines indicate $I_c^+$. The critical currents were determined using a voltage criterion of 5 μV. (d,e) Temperature dependence of $I_c^-$ and $I_c^+$ for the two bilayers. (f) Temperature dependence of η. Dashed vertical lines indicate $T_c$.

To place the critical currents on a physical scale, we compare the self-field data with a dirty-limit estimate of the depairing current based on the Clem-Kogan expression,[24] $I_{\mathrm{dep}}(0) = \frac{0.74 w {\Delta_0}^{3/2}}{e R_S \sqrt{\hbar D}}$, together with the approximate temperature dependence $I_{\mathrm{dep}}(T) \approx I_{\mathrm{dep}}(0)\left[1 - \left(\frac{T}{T_c}\right)^2\right]^{3/2}$. Here, $\Delta_0 = 1.76 k_B T_c$ is the weak-coupling BCS gap, $R_s = \frac{\rho_N}{t}$ is the sheet resistance, and $D = \frac{4 k_B}{\pi e |\mathrm{d}B_{c2}/\mathrm{d}T|}$ is the normal-state electron diffusion coefficient, with $B_{c2} = \mu_0 H_{c2}$. The zero-temperature dirty-limit penetration depth is estimated from $\lambda(0) = \sqrt{\frac{\hbar \rho_N}{\pi \mu_0 \Delta_0}}$. Using $\rho_N \approx 120$ μΩ·cm for the $Mo_2N$ single layers and $D$ in the range (3.5–3.8) × $10^{-5}$ $m^2/s$,[14] we obtain λ(0) ≈ 550 nm for $Mo_2N$(10)/FePt and approximately 450 nm for $Mo_2N$(20)/FePt, consistent with values reported for disordered molybdenum nitride.[25] For the nominal bridge geometry, these parameters yield zero-temperature depairing currents of approximately 3.6 mA and 12.4 mA for $Mo_2N$(10)/FePt and $Mo_2N$(20)/FePt, respectively. At the representative measurement temperatures, these estimates decrease to approximately 1.24 mA at 3.2 K and 4.93 mA at 4.4 K. The corresponding positive critical currents, approximately 0.43 mA and 2.2 mA, therefore reach about 35% and 45% of these nominal depairing estimates [**Figure** 4(d) and 4(e)]. These estimates assume a homogeneous superconducting layer of nominal thickness. Although the measured $T_c$ incorporates part of the superconducting suppression in the bilayer, the calculation does not explicitly resolve the proximity-induced variation of the superconducting properties near the $Mo_2N$/FePt interface. Representing this suppression by a reduced effective superconducting thickness would lower the estimated depairing currents and increase the corresponding current ratios, with the other parameters held fixed. Within this effective-thickness approximation, the corresponding fractions would therefore increase significantly above the nominal 35% and 45% values. By comparison, the opposite branches develop dissipation at much smaller fractions of this current scale.

**Figure 5** demonstrates that the diode response can be tuned and reversibly switched through magnetic control of the remanent configuration. For the angular measurements, the sample was first saturated in plane with a 0.5 T field applied at the selected angle, and the I–V curves were then measured at zero external field. Here, the angle denotes the in-plane preparation-field direction relative to the current. The response of $Mo_2N$(20)/FePt evolves systematically as the preparation field is rotated from parallel to the current at 0° toward perpendicular at 90° [**Figure** 5(a)]. The corresponding remanent stripe orientation was not independently imaged in the patterned bridge; the angular evolution is therefore interpreted in terms of the associated change in the remanent magnetic configuration. The separation between $I_c^+$ and $I_c^-$, and the corresponding efficiency, increases toward the 90° preparation configuration [**Figure** 5(b) and 5(c)]. The same angular trend is observed for $Mo_2N$(10)/FePt at 3.2 K, at lower absolute currents. This evolution is consistent with a change in the relation between the Lorentz force, the stripe channels, and their termination at the bridge edges. Both critical-current branches change with angle, indicating that the nonreciprocity involves the enhancement of the robust branch together with the reduction of the dissipative threshold.

We next examined whether the diode polarity can be reversed by rotating the remanent magnetization by 180°. For this purpose, the sample was first saturated in magnetic configuration I of **Figure** 3, and then small reverse-field excursions, $\mu_0\Delta H$, were successively applied in the opposite direction. After each excursion, the field was returned to zero and an I–V curve was measured at self-field. Representative curves are shown in **Figure** 5(d), where the diode response evolves continuously and eventually changes polarity. The corresponding evolution of $I_c^+$ and $I_c^-$ with $\mu_0\Delta H$ is presented in **Figure** 5(e). The diode state remains nearly unchanged for small reverse fields, up to approximately 0.01 T, and then switches within a field range consistent with the coercive-field range extracted from **Figure** 1(b). For reverse fields above about 0.04 T, the device exhibits a well-defined nonreciprocal response with opposite polarity. The same trend is captured by the efficiency data shown in **Figure** 5(f). These results indicate that, even after microfabrication, the diode response remains directly linked to the magnetization reversal of the FePt layer. The angular measurements demonstrate continuous tuning of the diode response through the in-plane preparation angle. The reverse-field excursions interchange the two critical-current branches, demonstrating magnetic control of the diode polarity at zero applied field during readout.

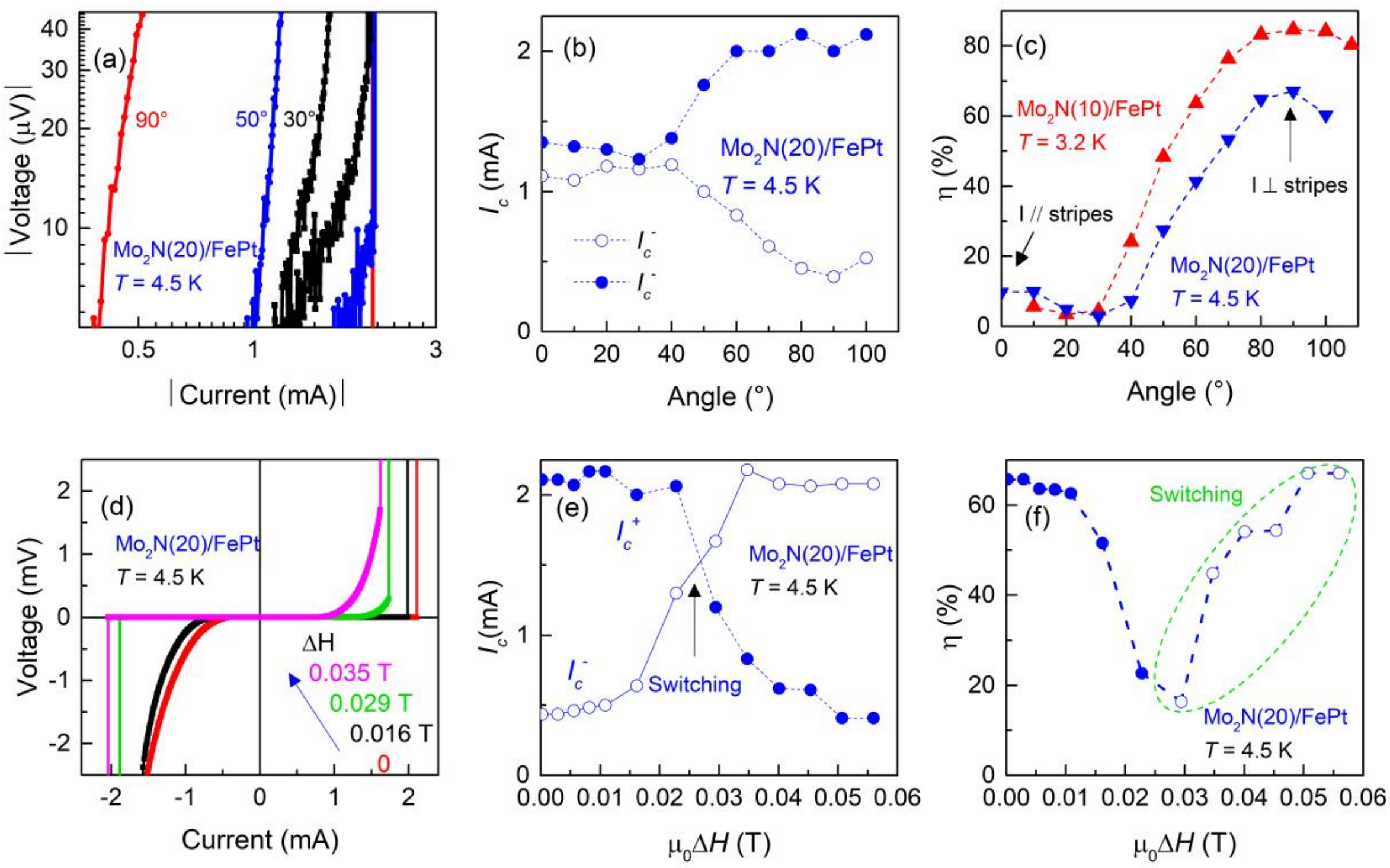

**Figure 5.** Magnetic tuning and reversal of the diode response. (a) |V| versus |I| for $Mo_2N$(20)/FePt at 4.5 K for different in-plane preparation-field angles relative to the current. (b) Angular dependence of $I_c^+$ and $I_c^-$. (c) Angular dependence of η for $Mo_2N$(10)/FePt at 3.2 K and $Mo_2N$(20)/FePt at 4.5 K. (d) Zero-field I-V curves for $Mo_2N$(20)/FePt after reverse-field excursions of amplitude $\mu_0\Delta H$. (e) $I_c^+$ and $I_c^-$ versus $\mu_0\Delta H$. (f) Magnitude of the diode efficiency versus $\mu_0\Delta$H; the polarity reversal is identified by the interchange of $I_c^+$ and $I_c^-$ in (e).

Our results share the qualitative phenomenology of the edge-controlled vortex diode described by Carapella *et al.*[6] for Py/Nb bilayers: strongly asymmetric thresholds for dissipative transport, reversal of the diode polarity with remanent magnetization, and suppression of nonreciprocity when the current becomes parallel to the magnetic stripes. In their system, the calculated interior stray field was approximately 27 mT, below the estimated Nb lower critical field of approximately 30 mT at 4.2 K, whereas the edge fields reached approximately 60 mT. These conditions supported preferential vortex and antivortex nucleation at opposite edges, followed by polarity-dependent penetration and guided motion along the stripes. Their simulations further showed that interior vortex–antivortex nucleation reduces the asymmetry, providing an explanation for its experimental disappearance below $T_c$ as the lower critical field decreases. In $Mo_2N$/FePt, however, $\eta$ increases over the measured temperature range, reaching approximately 83% at 6 K in the bilayer with $T_c \approx 6.5$ K. This difference indicates that the shared transport phenomenology need not imply identical conditions for vortex nucleation in the two systems. The magnetic and superconducting scales of $Mo_2N$/FePt differ substantially. From the estimated values of $\lambda(0)$ and $\xi(0)$,[14] $\mu_0 H_{c1}(0)$ is approximately 2.8 and 4.1 mT for the two bilayers, respectively, well below the calculated interior stripe fields of approximately 160–180 mT. These estimates provide an illustrative superconducting field scale rather than the perpendicular flux-entry threshold of the finite bridges. The alternating field has a period of approximately 110 nm, smaller than the penetration depth, and acts on a superconducting layer much thinner than that depth. Consequently, the existence of a stable interior vortex–antivortex population cannot be inferred from the local field amplitude alone. Bobba *et al.*[20] demonstrated that the occurrence of spontaneous vortex–antivortex populations in stripe-domain hybrids depends on the magnetic texture and superconducting dimensions. Their results support this general dependence, although their quantitative nucleation criteria do not apply directly to the thin-film, narrow-domain regime considered here, for which the nucleation threshold remains to be calculated. Despite the substantial interior fields, the positive-current branch remains below the experimental voltage criterion up to abrupt switching, while the opposite branch develops dissipation at much lower current. Flux entry at the finite bridge boundaries is distinct from spontaneous vortex–antivortex nucleation in the stripe interior: near the stripe terminations, the field is asymmetrically enhanced to approximately 280 mT and develops strong longitudinal gradients. This edge-field magnitude exceeds both the homogeneous lower-critical-field estimates and the characteristic field scale for suppressing the vortex-entry barrier under a uniform perpendicular field,[26] used here as qualitative reference scales; together with the contrasting current branches, these localized edge features are consistent with preferential edge injection and polarity-dependent vortex transport. Within this interpretation, magnetic confinement opposing inward penetration can favor a vortex-depleted transport region for the robust polarity, whereas the opposite polarity permits guided vortex and antivortex motion through the bridge. Transport alone, however, cannot distinguish such a depleted region from one containing immobilized vortices.

Moreover, FePt enables the qualitative diode phenomenology associated with stripe domains and finite magnetic edges using a ferromagnetic thickness of only 56 nm, compared with the 480–640 nm Py layers studied by Carapella *et al.*[6] This approximately tenfold reduction decreases the magnetic-layer thickness that must be removed during bridge patterning, while the entire bilayer is grown at room temperature. For comparable resistivities, a thinner metallic ferromagnet also reduces parallel normal-state conduction relative to a thicker magnetic layer, although FePt still dominates the normal-state conductance of the present bilayers. Compared with the previous field-free SC/FM hybrid diodes discussed above,[7,10-13] the efficiency of approximately 85%, achieved using a 56-nm-thick FePt layer and room-temperature growth, highlights the potential of this simple bilayer architecture for high-efficiency, magnetically programmable superconducting diodes.

## 3. Conclusion

In summary, we demonstrate a field-free superconducting diode effect in room-temperature-grown $Mo_2N$/FePt bilayers, with field-free diode efficiencies up to approximately 85%, continuous angular tunability, and reversible magnetic programming. Micromagnetic simulations establish the stripe-modulated stray-field landscape and its asymmetric edge corrections. Together with the transport measurements, these results support a picture in which magnetic control of flux entry and vortex motion produces widely separated critical currents for opposite bias directions. The same strategy may be applicable to other superconductors in regimes of weak vortex pinning.

## 4. Experimental Section

$Mo_2N$/FePt bilayers were grown through reactive sputtering at room temperature on a Si (100) substrate. The base pressure in the chamber was $4 \times 10^{-5}$ Pa. The $Mo_2N$ layer was grown in a 7% $N_2$ atmosphere at 0.66 Pa ($N_2$+Ar), with the substrate positioned 0.07 m above the Mo target (50 W). The FePt layer was grown in pure argon at 0.4 Pa using 20 W with the same target-to-substrate distance. The bilayers consisted of 10- and 20-nm-thick $Mo_2N$ layers capped with ≈ 56 nm of FePt. The thicknesses were selected to preserve superconductivity above liquid helium temperature in $Mo_2N$ and to stabilize stripe domains in FePt [14,15,17]. Hereafter, the samples are denoted as $Mo_2N$(10)/FePt and $Mo_2N$(20)/FePt.

X-ray diffraction data (XRD, Θ–2Θ configuration) and X-ray reflectivity (XRR) were obtained using a PANalytical Empyrean diffractometer at 40 kV and 30 mA with CuKα radiation and an angular resolution of 0.013°. The high-angle XRD pattern highlighting the FePt(111) reflection, together with the XRR curves and corresponding Parratt fits, is presented in **Figure** S1 of the Supporting Information. Surface morphology and magnetic domain structure were analyzed by atomic force microscopy (AFM) and magnetic force microscopy (MFM) using a Park AFM system (Model NX10). Electrical transport measurements were performed on an 80 μm long ($L$) × 5 μm wide ($w$) bridge using standard four-terminal transport. A typical optical image of the bridges is shown in **Figure** S2. Magnetic measurements were done with a commercial SQUID. Bridges were fabricated using optical lithography and argon ion milling. Current-voltage (I-V) curves were obtained with a Keithley Nanovoltmeter Model 2182A and a Keithley Current Source Model 6221 AC/DC in synchronized mode with a 0.1 ms pulse duration. The main transport features reported here were reproducibly observed in at least two independently fabricated devices for each $Mo_2N$ thickness.

## Author Contributions

N. Haberkorn and J. Kim conceived the project. N. Haberkorn prepared the samples and performed the XRD and magnetization measurements. Y. Lee and J. Kim performed the MFM measurements. J. Cruz, J. Zapata, and N. Haberkorn performed the electrical transport measurements. E. De Biasi performed the micromagnetic simulations. All authors discussed the results and contributed to writing and revising the manuscript. All authors approved the final version of the manuscript.

## Acknowledgments

Computational resources were provided by the HPC cluster of the Physics Department at Centro Atómico Bariloche (CNEA).

## Data Availability Statement

The data supporting the findings of this study are available from the corresponding authors upon request.

## Conflict of Interest

The authors declare no conflict of interest.

## Funding

This work was supported by CONICET (PIP 11220210100263CO) and by the Ministry of Science and ICT through the National Research Foundation of Korea (Grant Nos. RS-2023-00222408, RS-2024-00410027, and RS-2026-25472018).

## Supporting Information

## High-Efficiency Field-Free Superconducting Diodes in $Mo_2N$/FePt Bilayers

This Supporting Information provides complementary structural and device-level characterization of the $Mo_2N$/FePt bilayers discussed in the main text. **Figure** S1 shows the high-angle X-ray diffraction data around the FePt(111) reflection together with the corresponding X-ray reflectivity curves and Parratt fits used to confirm the layer thicknesses [S1]. **Figure** S2 presents a representative optical micrograph of the lithographically defined transport bridge used for the electrical measurements. The following sections describe the micromagnetic model, relaxation protocol, and magnetostatic reconstruction used to obtain **Figure** 2 of the main text.

### Structural characterization

The structural characterization is consistent with the intended bilayer design. In the high-angle XRD scan, the most prominent feature is the FePt(111) reflection, while the $Mo_2N$ contribution remains weak due to the reduced thickness and disorder of the superconducting layer [S2]. The XRR oscillations are well reproduced by Parratt fits, supporting the nominal layer thicknesses and the overall quality of the stack.

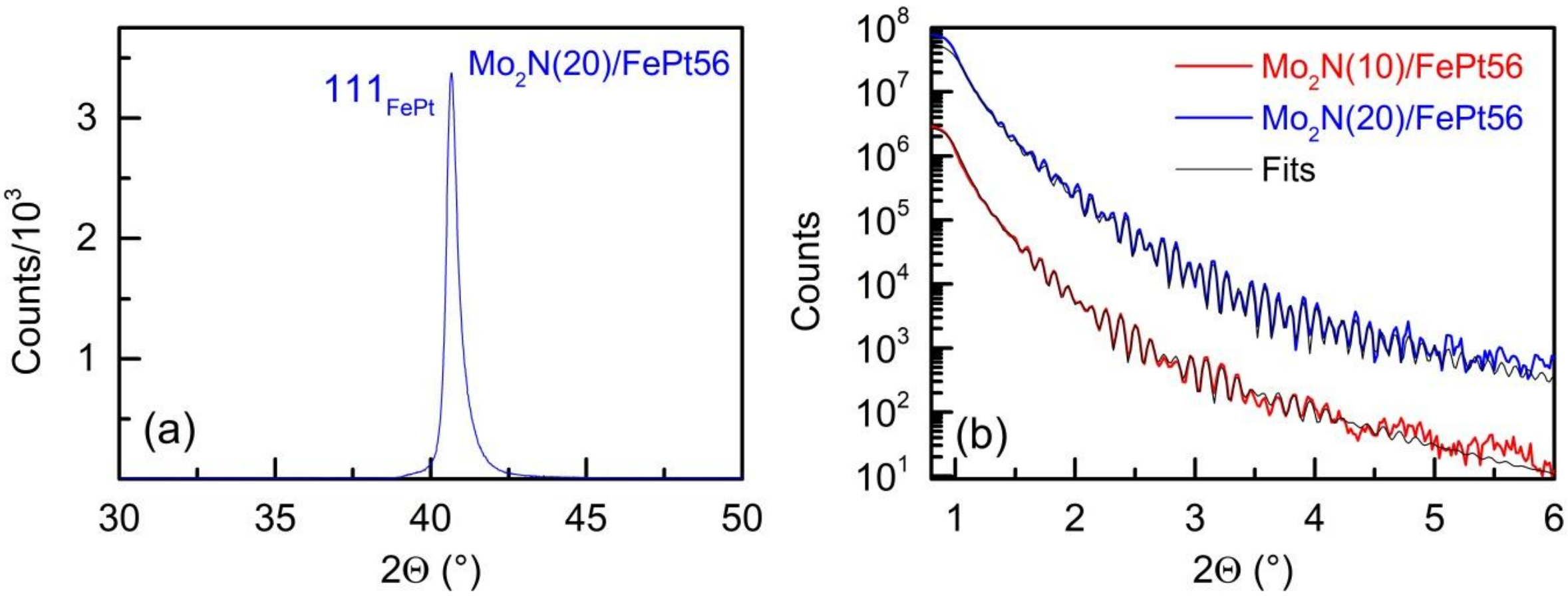

**Figure S1.** Structural characterization of the $Mo_2N$/FePt bilayers. (a) High-angle X-ray diffraction pattern measured in θ–2θ geometry for the $Mo_2N$(20)/FePt bilayer, showing the clear fcc FePt(111) reflection. The contribution from the $Mo_2N$ layer is much weaker because of its small thickness and disordered nature. (b) X-ray reflectivity curves for the $Mo_2N$(10)/FePt and $Mo_2N$(20)/FePt bilayers together with Parratt fits (black lines).[S1]

## Transport bridge geometry

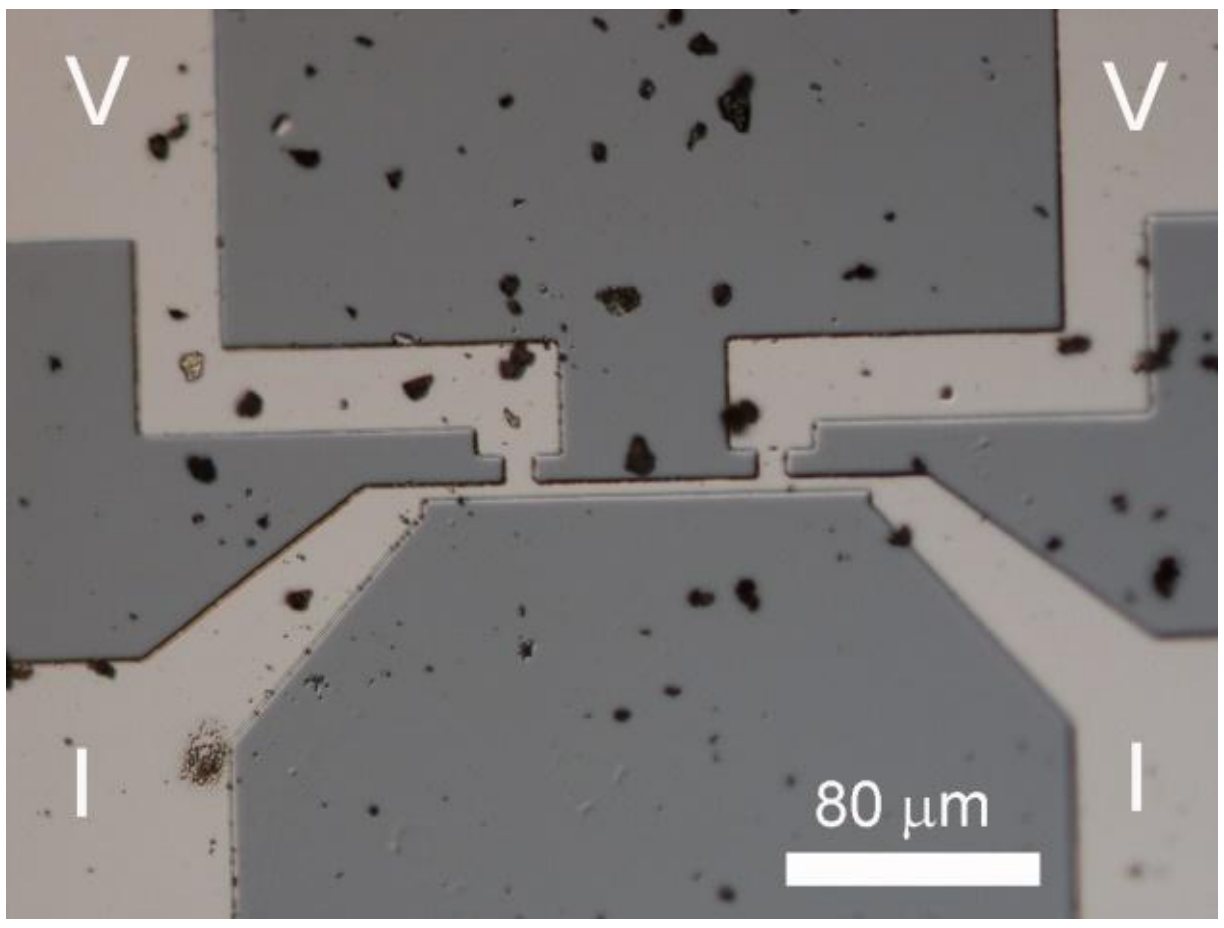

**Figure S2.** Optical image of a representative $Mo_2N$/FePt microbridge used for four-terminal transport measurements. The bridge was defined by optical lithography and argon ion milling. The active channel has dimensions of 80 μm in length and 5 μm in width. Current and voltage terminals are indicated in the image.

## Micromagnetic model and numerical discretization

The simulations were performed with MuMax3 version 3.10 [S3]. The saturation magnetization and perpendicular uniaxial anisotropy were chosen from the analysis of in-plane and out-of-plane magnetization measurements on FePt, complemented by the MFM observation of stripes with a characteristic period of approximately 110 nm. Previous studies of room-temperature-grown FePt stripe textures provide the material context [S4,S5]. The anisotropy easy axis was directed along z. The material and discretization parameters are collected in Table S1.

| **Parameter** | **Value** |
|---|---|
| Saturation magnetization $M_s$ | $1.08 \times 10^6$ A/m |
| Uniaxial anisotropy $K_{u1}$ | $1.9 \times 10^5$ J/m$^3$ |
| Anisotropy easy axis | z, perpendicular to the film |
| Exchange stiffness $A_{ex}$ | $1.20 \times 10^{-11}$ J/m |
| Cell dimensions | $4 \times 4 \times 3.5$ nm$^3$ |
| Grid dimensions | 1024 × 256 × 16 cells |
| Simulated dimensions | 4.096 µm × 1.024 µm × 56 nm |
| Periodic boundary condition | y, with 20 repetitions on each side |
| Boundary along x | Finite strip edges |
| Field evaluation plane | 10 nm below the lower FePt surface |
| Near-field cutoff radius | 50 nm from each source-cell center |
| Relaxed stripe period | 128 nm, eight periods along y |

Table S1. Parameters used for the micromagnetic relaxation and subsequent stray-field reconstruction.

The corresponding exchange length is

$$\ell_{\mathrm{ex}} = \sqrt{\frac{2A_{\mathrm{ex}}}{\mu_0 {M_s}^2}} \approx 4.05\, nm \quad \text{(S1)}$$

The cell dimensions are therefore comparable to the exchange length and smaller than the approximately 64 nm stripe half-period. The quality factor is $Q = \frac{2K_{\mathrm{u1}}}{\mu_0 {M_s}^2} \approx 0.26$, consistent with weak stripes in a film with perpendicular anisotropy and substantial in-plane remanence.

The *x* direction is parallel to the stripe axes. For the perpendicular-stripe transport configuration, current flows along *y*. Periodic boundary conditions along *y* reduce artificial end effects along the bridge length while retaining the finite edges across its width. The simulated width of 4.096 µm provides a representative finite strip for the 5 µm experimental bridge. The periodic cell allows only an integer number of stripe periods. The relaxed state contains eight periods over 1024 nm, giving $\Delta_S$ = 128 nm. The measured period of approximately 110 nm would correspond to about 9.3 periods in this cell, so the simulation reproduces the domain scale without exactly matching the measured period.

**Preparation of the remanent magnetic state**

The initial state was oriented along the in-plane saturation direction, with a 10% random perturbation added to seed domain nucleation. The system was first relaxed in a 0.8 T field along the same direction with damping parameter α = 1. The field was then reduced from 0.3 T to zero in 30 steps, with relaxation at each step using α = 0.5. Finally, two additional relaxations were carried out at 0.001 T and at zero field. The resulting remanent state was used for the field reconstruction. The large damping values were chosen for numerical relaxation and are not interpreted as physical dynamical parameters.

This staged relaxation follows the approach to remanence and permits the stripe pattern to develop as the field is reduced. It yields the remanent configuration shown in **Figure** 2(a), including the localized stripe distortions near one edge. In the convention adopted throughout the main text, the displayed state corresponds to remanence along -*x*, i.e., configuration I in **Figure** 3.

**Stray field reconstruction outside the ferromagnet**

An in-house C++ code reconstructed the perpendicular magnetostatic field from the relaxed magnetization. Each source cell was treated as a uniformly magnetized rectangular prism, with $M_j = M_s \hat{m}_j$. The evaluation plane was placed 10 nm below the lower FePt surface. Its distance from the center of the nearest cell layer is therefore 10 + 3.5/2 = 11.75 nm. This cell-center distance is distinct from the physical separation from the magnetic surface.

The total field is the sum of the cell contributions,

$$B_z(r_e) = \sum_j K_j(r_e) \cdot M_j \quad \text{(S2)}$$

where $K_j$ is the field kernel of source cell j. For source-center distance R ≤ 50 nm, the rectangular-prism field was evaluated analytically; for R > 50 nm, a point-dipole approximation was used. The closest cells consequently lie within the analytical regime. This hybrid calculation retains the finite-cell geometry near the observation point and uses the simpler far-field kernel for distant sources.

The field is expressed as $B_z$ or $\mu_0 H_z$ in tesla, or millitesla in the figure. In the exterior magnetostatic calculation, $B_z = \mu_0 H_z$. Superconducting screening and vortex nucleation are not included, so this field describes the magnetic source acting on the superconducting layer.

**Analytical field of a nearby rectangular cell**

Let the source-cell center be $r_m = (x_m, y_m, z_m)$, with dimensions $L_x \times L_y \times L_z$, and the observation point be $r_e = (x_e, y_e, z_e)$. The x faces are $x_i = x_m - L_x/2$ and $x_F = x_m + L_x/2$; the y and z faces are defined analogously. For each corner $(x', y', z')$, define

$$\Delta x = x_e - x', \ \Delta y = y_e - y', \ \Delta z = z_e - z' \quad \text{(S3)}$$

$$\tilde{r} = \sqrt{\Delta x^2 + \Delta y^2 + \Delta z^2} \quad \text{(S4)}$$

with $x'$ taking the values $x_i$ and $x_F$, and similarly for $y'$ and $z'$. Define $\delta_x F = F(x_F) - F(x_i)$, and the corresponding operators $\delta_y$ and $\delta_z$. The combined operator $\delta_x \delta_y \delta_z$ evaluates the expression at the eight corners with alternating signs: +1 for an even number of lower-face coordinates and −1 for an odd number. Rectangular-cell magnetostatic expressions are discussed in Ref. [S6]. With this convention, the cell contribution is

$$B_z = K_x M_x + K_y M_y + K_z M_z \quad \text{(S5)}$$

$$K_x = -\frac{\mu_0}{4\pi} \delta_x \delta_y \delta_z \operatorname{asinh}\left(\frac{\Delta y}{\sqrt{\Delta x^2 + \Delta z^2}}\right) \quad \text{(S6)}$$

$$K_y = -\frac{\mu_0}{4\pi} \delta_x \delta_y \delta_z \operatorname{asinh}\left(\frac{\Delta x}{\sqrt{\Delta y^2 + \Delta z^2}}\right) \quad \text{(S7)}$$

$$K_z = \frac{\mu_0}{4\pi} \delta_x \delta_y \delta_z \arctan\left(\frac{\Delta x \Delta y}{\Delta z \tilde{r}}\right) \quad \text{(S8)}$$

Here $M_x$, $M_y$, and $M_z$ are the components of the cell magnetization in A/m. The cell volume is already included through the integration over the prism and therefore does not multiply these exact kernels separately.

**Point dipole field of a distant cell**

For a distant source cell, let R = $r_e - r_m$ = (X,Y,Z), R = |R|, and V = $L_x \times L_y \times L_z$. The magnetic dipole moment is μ = VM, where M is the magnetization. The perpendicular dipole field is [S7]

$$B_{z,\mathrm{dip}} = \frac{\mu_0 V}{4\pi}\left[\frac{3Z(M\cdot R)}{R^5} - \frac{M_z}{R^3}\right] \quad \text{(S9)}$$

Equivalently, $B_z$,dip = a·M, with

$$a_x = \frac{\mu_0 V}{4\pi}\frac{3ZX}{R^5} \quad \text{(S10)}$$

$$a_y = \frac{\mu_0 V}{4\pi}\frac{3ZY}{R^5} \quad \text{(S11)}$$

$$a_z = \frac{\mu_0 V}{4\pi}\left[\frac{3Z^2}{R^5} - \frac{1}{R^3}\right] \quad \text{(S12)}$$

The volume V is explicit in these far-field kernels. Using magnetization, rather than magnetic moment, in both the exact and dipole expressions gives a consistent normalization for the summed field.

### Spatial interpretation of the simulated profiles

The relaxed texture consists of alternating perpendicular magnetization components with stripe axes along x. Away from the localized distorted region, the stripes become approximately straight and parallel. The transverse profile in **Figure** 2(b) samples the periodic field modulation, while the two paths in **Figure** 2(c) follow opposite stripe polarities. Their interior fields retain their signs and approach magnitudes of approximately 160 mT. In the coordinate convention displayed in **Figure** 2, the positive-field path is enhanced near the right edge and reduced near the left edge, while the negative-field path shows the complementary behavior, with its magnitude enhanced near the left edge.

A transverse Fourier component of a planar magnetostatic field with wave number $k_n$ decays outside the magnetic surface as

$$B_n(d) = B_n(0)e^{-|k_n|d}, \; k_n = \frac{2\pi n}{p} \quad \text{(S13)}$$

where $d$ is the distance from that surface. Higher spatial harmonics therefore attenuate faster than the fundamental, contributing to the smooth, nearly sinusoidal field profile. For $d$ = 10 nm and $p$ = 128 nm, the third harmonic receives a relative attenuation factor $\exp\left(\frac{-4\pi d}{p}\right) \approx$ 0.37 compared with the fundamental. This factor describes propagation alone; the actual harmonic ratio also depends on the magnetization texture and the film thickness. Here $d$ is the 10 nm surface separation, while the distances from individual cell centers enter the discrete magnetostatic kernels.

The strongest $\partial B_z/\partial x$ peaks are localized near the two finite $x$ edges, reaching approximately 8 mT/nm. Weaker longitudinal variations are also visible near the stripe distortions, so the field gradient is not strictly confined to the edges. The in-plane remanence produces opposite surface magnetic charges, $\sigma = M \cdot n$, on the faces normal to $x$. Their field adds opposite-signed corrections near the two ends to the transverse stripe modulation, providing a magnetostatic origin for the edge asymmetry. The positions of vortex nucleation and the associated forces depend on the superconducting response to this field.

**Supporting Information References**